\documentclass[11pt]{article}
\newcommand{\vev}[1]{\left<{#1}\right>}

\newcommand{\cket}[1]{\left|{#1}\right>_{\rm \!c}}

\newcommand{\iket}[1]{|{#1}\rangle\!\rangle}

\newcommand{\tfrac}[2]{{\textstyle\frac{#1}{#2}}}

\newcommand{\bS}{{\bf S}}
\newcommand{\bG}{{\bf G}}
\newcommand{\IIxI}[2]{\left[\!\!\begin{array}{c}#1\\#2\end{array}\!\!\right]}
\newcommand{\IIxII}[4]{\left[\!\!\begin{array}{cc}#1&#2\\#3&#4\end{array}\!\!\right]}
\begin{document}

\renewcommand{\thefootnote}{\fnsymbol{footnote}}
\baselineskip = 0.6cm
\pagestyle{plain}

\thispagestyle{empty}
\setcounter{page}{0}

\baselineskip 5mm
\hfill\vbox{\hbox{YITP-01-56}
            \hbox{hep-th/0108093} }

\baselineskip0.8cm\vskip2cm

\begin{center}
 {\large\bf Bulk-Boundary Propagator in Liouville Theory on a Disc}
\end{center}

\vskip10mm

\baselineskip0.6cm
\begin{center}
  Kazuo Hosomichi\footnote{
 \tt hosomiti@yukawa.kyoto-u.ac.jp}\\         \vskip2mm
{\it Yukawa Institute for Theoretical Physics \\
     Kyoto University, Kyoto 606-8502, Japan} \vskip3mm
\end{center}

\vskip8mm\baselineskip=3.5ex
\begin{center}{\bf Abstract}\end{center}\par\smallskip

   We study Liouville theory on worldsheets with boundary using
 the solutions of Knizhnik-Zamolodchikov equation involving
 a degenerate representation of the Virasoro algebra.
   The expression for bulk-boundary propagator on a disc
 is proposed.

\vspace*{\fill}
\noindent August~~2001


\newpage

\renewcommand{\thefootnote}{\arabic{footnote}}
\setcounter{footnote}{0}
\setcounter{section}{0}
\baselineskip = 0.6cm
\pagestyle{plain}
\renewcommand{\thesection}{\arabic{section}.}
\renewcommand{\thesubsection}{\arabic{section}.\arabic{subsection}.}
\renewcommand{\thefootnote}{\arabic{footnote}}
\setcounter{footnote}{0}


   Liouville theory is an interacting two-dimensional
 conformal field theory that has many applications in string theory.
 It was originally studied as a theory of two-dimensional gravity,
 and recently it has also applied to the study of strings propagating
 near the singularities of manifolds.
 It plays a key role in understanding the holography in string theory,
 since most of the known backgrounds realizing the holography
 have a radial direction which is described by a linear-dilaton like
 theory.
 To study the property of D-branes on such backgrounds from the CFT
 viewpoint, it is therefore necessary to understand the Liouville
 theory on open worldsheets.

   Liouville theory on open worldsheets was analyzed in \cite{FZZ, ZZ}
 where some basic correlators were derived.
 They were based on bootstraps and free field techniques which were
 first developed in \cite{DO,ZZ2,T} for the case without boundary.
 See also \cite{PT,T2,T3}.
 Our analysis in this letter is along these path.

~

   Liouville theory on a worldsheet with boundary is defined by
 the action
\begin{eqnarray}
  I&=&\frac{1}{8\pi}\int_\Sigma d^2\sigma \sqrt{g}
     \left[g^{mn}\partial_m\phi\partial_n\phi
    +\sqrt{2}QR\phi+8\pi\mu e^{\sqrt{2}b\phi}\right]
  \nonumber \\ &&
    +\frac{1}{4\pi}\int_{\partial\Sigma}d\xi g^{1/4}
     \left[\sqrt{2}QK\phi+4\pi\mu_B e^{b\phi/\sqrt{2}}\right] ,
\end{eqnarray}
 where $K$ is the curvature of the boundary and $\mu_B$ is
 called the boundary cosmological constant.
 $\mu_B$ can take different values for each connected component of
 the boundary and, what is more, $\mu_B$ can be different for
 each boundary segment bounded by two boundary vertex operators.

   When $\mu$ and $\mu_B$ vanish, the theory reduces to a free theory
 with the stress tensor
\begin{equation}
  T = -\frac{1}{2}(\partial\phi\partial\phi-\sqrt{2}Q\partial^2\phi)
\end{equation}
 and the center $c=1+6Q^2$.

   We have two kinds of primary fields in Liouville theory with boundary.
 Bulk primaries $V_\alpha(z)=e^{\sqrt{2}\alpha\phi(z)}$ correspond
 to closed string modes while boundary primaries
 $B_\alpha(x)=e^{\alpha\phi(x)/\sqrt{2}}$ correspond to open string modes.
 They are of conformal weight $h_\alpha=\alpha(Q-\alpha)$.
 We mainly use coordinates $z,w,\ldots$ for positions of bulk fields
 and $x,y,\cdots$ for boundary fields.

   As was studied in \cite{FZZ,ZZ}, it is natural to regard $\mu_B$
 as labeling D-branes or Cardy states in Liouville theory.
 The classification of Cardy states has been done in \cite{ZZ} using
 the modular property of Virasoro characters.
 Consistent Cardy states are expressed as linear combination of
 Ishibashi states, and the coefficients are proportional to
 the one-point function of a bulk field $\vev{V_\alpha(z)}$ on a disc.
 This was obtained in \cite{FZZ}:
\begin{eqnarray}
  \vev{V_\alpha(z)}_{\mu_B} &=& U(\alpha;s)|z-\bar{z}|^{-2h_\alpha},
 \nonumber \\
  U(\alpha;s) &=& 2 \cosh\left[2\pi s(2\alpha-Q)\right]
  \{\mu\pi\gamma(b^2)\}^{(Q-2\alpha)/2b}
  (2\alpha-Q)\Gamma[(2\alpha-Q)b]\Gamma[(2\alpha-Q)/b],
\label{U}
\end{eqnarray}
 where $\gamma(x)=\Gamma(x)/\Gamma(1-x)$ and we introduced
 a new parameter $s$ for labeling the Cardy states $\cket{s}$.
 It is related to $\mu_B$ via
\begin{equation}
  \mu_B = \cosh(2\pi sb)\left(\frac{\mu}{\sin\pi b^2}\right)^{1/2}.
\label{bcc-p}
\end{equation}

   Our goal is to obtain the expression for bulk-boundary propagator
 $\vev{V_\alpha(z)B_\beta(x)}_s$ on a disc.
 What we have to determine is the structure constant
 $R(\alpha,\beta;s)$ defined by
\begin{equation}
  \vev{V_\alpha(z)B_\beta(x)}_{s}=
  |z-\bar{z}|^{h_\beta-2h_\alpha}|z-x|^{-2h_\beta}R(\alpha,\beta;s).
\end{equation}
 To obtain this we use the ($1+2$)-point function
 with an auxiliary insertion of a degenerate operator $B_{-b/2}(y)$.
 Conformal invariance restricts the correlator to take the form
\begin{eqnarray}
\lefteqn{
  \vev{V_\alpha(z)B_\beta(x)B_{-b/2}(y)}^{\rm (upper~half~plane)}_{s_L,s_R}
} \nonumber \\
 &=& (x-z)^{-h_\beta-h_{-b/2}}(x-\bar{z})^{-h_\beta+h_{-b/2}}
     |z-\bar{z}|^{h_\beta-2h_\alpha+h_{-b/2}}(y-\bar{z})^{-2h_{-b/2}}
  \nonumber \\
 & &\times \vev{V_\alpha(0)B_\beta(1)B_{-b/2}(\eta)}^{\rm (disc)}_{s_L,s_R},
  ~~~~\eta=\tfrac{(y-z)(x-\bar{z})}{(y-\bar{z})(x-z)}
\end{eqnarray}
 This correlator has two boundary segments since there are two
 boundary operator insertions.
 The suffices $L,R$ mean the left or right of $B_{-b/2}$ on the real axis.
 The two boundary cosmological constants must satisfy
 $s_L\pm s_R = \pm' ib/2$ so as to be consistent with
 the fusion rule\cite{ZZ}.
 Otherwise the null vector actually does not decouple.

   The $(1+2)$-point function can be expressed in a number of ways:
\begin{eqnarray}
\lefteqn{
  \vev{V_\alpha(0)B_\beta(1)B_{-b/2}(\eta)}^{\rm (disc)}_{s_L,s_R}
}\nonumber \\
 &=& c_+(\beta;s_L,s_R)R(\alpha,\beta-\tfrac{b}{2};s_R)
     G_{\alpha,\beta}(\eta)e^{i\pi b\beta/2}
 \nonumber \\ && \!\!\!\!\!\!
   +~ c_-(\beta;s_L,s_R)R(\alpha,\beta+\tfrac{b}{2};s_R)
      G_{\alpha,Q-\beta}(\eta)e^{i\pi b(Q-\beta)/2}
 \nonumber \\
 &=& c_+(\beta;s_R,s_L)R(\alpha,\beta-\tfrac{b}{2};s_L)
     G_{\alpha,\beta}(\eta e^{-2\pi i})e^{-i\pi b\beta/2}
 \nonumber \\ &&\!\!\!\!\!\!
  +~ c_-(\beta;s_R,s_L)R(\alpha,\beta+\tfrac{b}{2};s_L)
     G_{\alpha,Q-\beta}(\eta e^{-2\pi i})e^{-i\pi b(Q-\beta)/2},
\end{eqnarray}
 where the equality should hold at least up to overall factors.
 Here $G_\beta(\eta)$ is the following solution of KZ equation:
\begin{equation}
  G_{\alpha,\beta}(\eta) =\eta^{b\alpha}(1-\eta)^{b\beta}
   F\left(b(2\alpha+\beta-Q-\tfrac{b}{2}), b(\beta-\tfrac{b}{2}),
          b(2\beta-b);1-\eta\right).
\end{equation}
 We assume that the correlator is analytic on the complex
 $\eta$-plane except on the positive half of the real axis.
 The expression $\eta e^{-2\pi i}$ indicates going clockwise
 around $\eta=0$ once.
 $c_\pm(\beta;s_L,s_R)$ are the OPE coefficients:
\begin{equation}
  B_\beta(x)B_{-b/2}(y)_{s_L,s_R} \stackrel{x<y}{\sim}
  |y-x|^{b\beta}c_+(\beta;s_L,s_R)B_{\beta-b/2}(x)
 + |y-x|^{b(Q-\beta)}c_-(\beta;s_L,s_R)B_{\beta+b/2}(x)
\end{equation}
 They are calculated as the free field correlators
 with an appropriate insertion of boundary screening operator
 $S_B = \int_{\partial\Sigma}B_b$:
\begin{eqnarray}
  c_+(\beta;s_L,s_R)
  &=& \vev{B_\beta B_{-b/2}B_{Q-\beta+b/2}}_{\rm free}
  ~=~ 1,
 \nonumber \\
  c_-(\beta;s_L,s_R)
  &=& \vev{(-\mu_BS_B)B_\beta B_{-b/2}B_{Q-\beta+b/2}}_{\rm free}
 \nonumber \\
  &=& \frac{2b^2}{\pi}\Gamma(1-2b\beta)\Gamma(2b\beta-bQ)
      \cos\pi(b\beta-\tfrac{bQ}{2})(\mu\pi\gamma(b^2))^{1/2}
      \cos\pi(b\beta-\tfrac{bQ}{2}\mp 2ibs_R)
 \nonumber \\
  & & ~~~~(s_R = s_L \pm \tfrac{ib}{2}).
\end{eqnarray}

   We can derive some recursion relations between $R(\alpha,\beta;s)$
 from the transformation property and monodromy of
 hypergeometric functions.
 We shall do this by rewriting the $(1+2)$-point function as a linear
 sum of the solutions $H_{\alpha,\beta}(\eta)$
 and $H_{Q-\alpha,\beta}(\eta)$ that diagonalize the monodromy
 around $\eta=0$.
 They are defined as
\begin{equation}
  H_{\alpha,\beta}(\eta)=\eta^{b\alpha}(1-\eta)^{b\beta}
  F\left(b(2\alpha+\beta-Q-\tfrac{b}{2}),b(\beta-\tfrac{b}{2}),
         b(2\alpha-b);\eta\right)
\end{equation}
 and are related to $G_{\alpha,\beta}$ via
\begin{eqnarray}
    (G_{\alpha,\beta},G_{\alpha,Q-\beta})
 &=&(H_{\alpha,\beta},H_{Q-\alpha,\beta})
\IIxII{x_{\beta,\alpha}}{x_{Q-\beta,\alpha}}
      {x_{\beta,Q-\alpha}}{x_{Q-\beta,Q-\alpha}}, \nonumber \\
  x_{\beta,\alpha} &=&
    \frac{\Gamma(2b\beta-b^2)\Gamma(1+b^2-2b\alpha)}
         {\Gamma(1+\frac{b^2}{2}-2b\alpha+b\beta)
          \Gamma(b\beta-\frac{b^2}{2})}.
\end{eqnarray}
 There are subtle phase factors arising in deriving
 the recursion relation, which should be handled carefully
 so that the resultant relation is symmetric under
 the exchange of $s_L$ and $s_R$.
 We must also pay attention to the relation
\begin{equation}
  \vev{V_\alpha(0)B_\beta(1)B_{-b/2}(\eta)}^{\rm (disc)}_{s_L,s_R}
 =\eta^{-2h_{-b/2}}
  \vev{V_\alpha(0)B_\beta(1)B_{-b/2}(\eta^{-1})}^{\rm (disc)}_{s_R,s_L},
\end{equation}
 which merely corresponds to flipping the real axis
 of the upper half-plane.
 We finally find the following relation:
\begin{equation}
  iX^+\IIxI{c_+(\beta;s_R,s_L)R(\alpha,\beta-\tfrac{b}{2};s_L)}
           {c_-(\beta;s_R,s_L)R(\alpha,\beta+\tfrac{b}{2};s_L)}
 = X^-\IIxI{c_+(\beta;s_L,s_R)R(\alpha,\beta-\tfrac{b}{2};s_R)}
           {c_-(\beta;s_L,s_R)R(\alpha,\beta+\tfrac{b}{2};s_R)}
\label{recur}
\end{equation}
 where the components of $X^\pm$ are given by
\begin{equation}
  x^\pm_{\beta,\alpha}=
  e^{\pm i\pi b(\alpha+\frac{\beta}{2}-\frac{3Q}{4})}x_{\beta,\alpha}.
\end{equation}
 The phases are fixed so that we can find a matrix $Y$ satisfying
 $YX^\mp = \pm iX^\pm$, ensuring that the relation is symmetric
 under the exchange of $s_L$ and $s_R$.

   Let us pose a while and see the consistency of the relation
 (\ref{recur}) with the reflection symmetry of the vertices.
 The structure constant $R(\alpha,\beta;s)$ should be invariant
 under the reflection of the bulk vertex operator
 $V_\alpha \Leftrightarrow V_{Q-\alpha}$:
\begin{equation}
 V_\alpha=D(\alpha)V_{Q-\alpha},~~~
 D(\alpha)= -[\mu\pi\gamma(b^2)]^{(Q-2\alpha)/b}
   \frac{\Gamma((2\alpha-Q)b)\Gamma((2\alpha-Q)/b)}
        {\Gamma(-(2\alpha-Q)b)\Gamma(-(2\alpha-Q)/b)}.
\end{equation}
 and the reflection of the boundary operator
 $B_\beta \Leftrightarrow B_{Q-\beta}$:
\begin{eqnarray}
 B_\beta &=& d(\beta;s,s) B_{Q-\beta},\nonumber \\
 d(\beta;s_L,s_R) &=&
 \frac{\bG(Q-2\beta)\bG(2\beta-Q)^{-1}
       \{\mu\pi\gamma(b^2)b^{2-2b^2}\}^{(Q-2\beta)/2b}}
 {\bS(\beta+is_L+is_R)\bS(\beta-is_L-is_R)
  \bS(\beta+is_L-is_R)\bS(\beta-is_L+is_R)}.
\label{bref}
\end{eqnarray}
 Here we used some of the functions $\Upsilon$, $\bS$,
 $\bG$ introduced in \cite{DO, ZZ2, FZZ}.
 They are characterized by the shift relations
\begin{equation}
  \Upsilon(x+b)=b^{1-2bx}\gamma(bx)\Upsilon(x),~~~~
  \bS(x+b)=2\sin(\pi bx)\bS(x), ~~~~
  \bG(x+b)=\frac{b^{\frac{1}{2}-bx}}{\sqrt{2\pi}}\Gamma(bx)\bG(x),
\end{equation}
 and those with $b$ replaced with $1/b$.
 We can check that if we can find a solution $R(\alpha,\beta;s)$ of
 the recursion relation, then its reflections
\begin{equation}
  D(\alpha)R(Q-\alpha,\beta;s),~~~
  d(\beta;s,s)R(\alpha,Q-\beta;s)
\end{equation}
 are also solutions of the same recursion relation.

    By inserting $s_R=s_L+\tfrac{ib}{2}$ into (\ref{recur})
 we obtain partial difference equations for $R(\alpha,\beta;s)$.
 To write down the solution, it seems that the most efficient way
 is to take the Fourier transform:
\begin{equation}
  \tilde{R}(\alpha,\beta;p)
  = \tfrac{1}{2}\int_{-\infty}^{\infty} ds e^{4\pi sp}R(\alpha,\beta;s),~~~~
  R(\alpha,\beta,s)
  = -i\int_{-i\infty}^{i\infty}dp e^{-4\pi sp}\tilde{R}(\alpha,\beta,p),
\end{equation}
 in terms of which the difference equation becomes simply
\begin{eqnarray}
&&
  \tilde{R}(\alpha,\beta;p+b)
       \sin\pi b(p+\alpha-\tfrac{\beta}{2}+\tfrac{Q}{2})
       \sin\pi b(p-\alpha-\tfrac{\beta}{2}+\tfrac{3Q}{2})
\nonumber \\ &=&
  \tilde{R}(\alpha,\beta;p)
       \sin\pi b(p+\alpha+\tfrac{\beta}{2}-\tfrac{Q}{2})
       \sin\pi b(p-\alpha+\tfrac{\beta}{2}+\tfrac{Q}{2})
\nonumber \\ &=&
  \tilde{R}(\alpha,\beta+b;p+\tfrac{b}{2})
  \frac{\pi^2 b^2(\mu\pi\gamma(b^2))^{1/2}
        \Gamma(1-2b\beta)\Gamma(1-b^2-2b\beta)}
       {\Gamma(b\beta)\Gamma(1-b\beta)^3
        \Gamma(1-b\beta-2b\alpha+bQ)\Gamma(1-b\beta+2b\alpha-bQ)}.
\end{eqnarray}
 A solution is given by
\begin{eqnarray}
  \tilde{R}(\alpha,\beta;p) &=&
  2\pi(\mu\pi\gamma(b^2)b^{2-2b^2})^{(Q-2\alpha-\beta)/2b}
  \frac{\bG(Q)\bG(\beta)\bG(Q-2\beta)\Upsilon(2\alpha)}
       {\bG(Q-\beta)^3\bG(2Q-2\alpha-\beta)\bG(2\alpha-\beta)}
  \nonumber \\ && \times 
  \bS( p+\tfrac{\beta+2\alpha-Q}{2})
  \bS( p+\tfrac{\beta-2\alpha+Q}{2})
  \bS(-p+\tfrac{\beta+2\alpha-Q}{2})
  \bS(-p+\tfrac{\beta-2\alpha+Q}{2}).
\label{Rtil}
\end{eqnarray}
 The normalization was fixed so that we have $R(\alpha,0;s)=U(\alpha;s)$.

   We have to analyze the transformation property of the above
 solution under the reflection (\ref{bref}).
 To do this we check whether or not the value of $R(\alpha,\beta;s)$
 at $\beta=Q$ is as required from the reflection symmetry:
\begin{equation}
  d(\beta;s,s)\times(-i)\int dp e^{-4\pi sp}\tilde{R}(\alpha,Q-\beta;p)
  \stackrel{\beta\to 0}{\rightarrow} U(\alpha;s).
\end{equation}
 In doing this we have to be careful for the fact that when a factor
 like $e^{4\pi sp_0}$ is multiplied onto the inverse Fourier transform
 ($p$-integration), it shifts the contour of $p$-integration
 and possibly picks up some poles.
 A calculation shows that our structure constant satisfies
 the above equation, ensuring the reflection symmetry at $\beta=0$.

   As we have seen above, the solution was obtained by solving the
 recursion relation (\ref{recur}) and imposing the reflection symmetry.
 The normalization was fixed by matching with $U(\alpha;s)$ in the limit
 $\beta\rightarrow 0$.
 Any solution satisfying these conditions should be equal to the above
 $R(\alpha,\beta;s)$ for all $\beta\in\{b{\bf Z}+b^{-1}{\bf Z}\}$
 due to recursion relation (\ref{recur}) and that associated with
 the degenerate field $B_{-1/2b}$.
 In this sense the solution is unique.

   It is known from the perturbative analysis based on
 path-integral formalism that the correlators diverge when the
 non-conservation of Liouville momentum can be screened by a
 non-negative integer number of screening operators.
 Our $R(\alpha,\beta;s)$ indeed has the corresponding
 poles at $2\alpha+\beta=Q-nb$.
 For the first few of them we can also check that the residue
 is precisely equal to the free-field correlators as expected.

~
 
   If a bulk operator $V_\alpha$ approaches the boundary, it can be
 expanded as a sum over boundary fields $B_\beta$.
 One naively expects that the expansion obeys the following formula
\begin{eqnarray}
  V_{\alpha}(z)&\stackrel{z\rightarrow x}{\sim}&
  \tfrac{1}{2} \int d\beta
  |z-\bar{z}|^{h_\beta-2h_\alpha} R(\alpha,\beta;s)B_{Q-\beta}(x)
    \nonumber \\ &=&
  \tfrac{1}{2i}\int d\beta dp
  |z-\bar{z}|^{h_\beta-2h_\alpha} e^{-4\pi sp}
  \tilde{R}(\alpha,\beta;p)B_{Q-\beta}(x).
\label{VBexp}
\end{eqnarray}
 where the $\beta$-integration should be done over 
 $\beta\in \frac{Q}{2}+i{\bf R}$~ for $\alpha\in \frac{Q}{2}+i{\bf R}$,
 and suitably deformed for generic $\alpha$ in order to
 ensure the analyticity.
 Similar deformation of contour is also assumed for $p$-integration.
 As a special case, take as $\alpha$ those values corresponding to
 degenerate representation of the Virasoro algebra:
\begin{equation}
  2\alpha_{k,l}=Q-kb-lb^{-1},~~~(k,l\in {\bf Z}_{\ge 1}).
\end{equation}
 We then expect that $V_\alpha$ is expanded into boundary degenerate
 operators $B_{\beta_{m,n}}$ with
\begin{equation}
  2\beta_{m,n}=Q-mb-nb^{-1},~~~
 (m = 1,3,\cdots, 2k-1,~~~
  n = 1,3,\cdots, 2l-1).
\label{degB}
\end{equation}
 Our $R(\alpha,\beta;s)$ agrees with this expectation.
 To see this, note first that the integral over $\beta$ and $p$
 has only contribution from poles due to the vanishing factor
 $\Upsilon(2\alpha)$ in the integrand.
 In order to cancel this factor, we actually need
 the degeneration of two poles that pinch the integration contour.
 By analyzing the location and the degeneracy of the poles of
\begin{eqnarray}
\lefteqn{  \tilde{R}(\alpha,\beta;s)d(Q-\beta;s,s) } \nonumber \\
 &\sim& \int dp
            \frac{\bG(2\beta-Q)\Upsilon(2\alpha)
            \bS( p+\tfrac{\beta+2\alpha-Q}{2})
            \bS( p+\tfrac{\beta-2\alpha+Q}{2})
            \bS(-p+\tfrac{\beta+2\alpha-Q}{2})
            \bS(-p+\tfrac{\beta-2\alpha+Q}{2})}
           {\bG(\beta)\bG(Q-\beta)\bG(2Q-2\alpha-\beta)\bG(2\alpha-\beta)},
\end{eqnarray}
 we find that the degenerate bulk field $V_{\alpha_{k,l}}$
 is expanded into degenerate boundary fields $B_{\beta_{m,n}}$
 with $(m,n)$ precisely given in (\ref{degB}).
 Note that the reflection coefficient $d(\beta;s,s)$ becomes singular
 when $\beta$ belongs to a degenerate representation.
 Therefore, only $B_{\beta_{m,n}}$ (and not their reflection transform)
 can appear with finite coefficient in the expansion of $V_{\alpha_{k,l}}$.

~

   The Fourier transformed structure constant $\tilde{R}(\alpha,\beta;p)$
 may well be thought of as the fusion coefficient of an Ishibashi state
 $\iket{p}$ with a boundary primary $B_\beta$.
 It seems therefore reasonable that the bulk-boundary structure constant
 becomes simpler if we take Ishibashi states rather than Cardy states.
 In \cite{ZZ} the Cardy states $\cket{m,n}$ corresponding to degenerate
 representations of Virasoro algebra have also been constructed
 and analyzed in some detail.
 From the comparison of the wave functions for Cardy states
 $\cket{s}$ and $\cket{m,n}$, it is expected that the bulk-boundary
 structure constant for degenerate Cardy states might be given by
\begin{equation}
  R(\alpha,\beta)_{m,n} = 2i\int_{-i\infty}^{i\infty}dp
  \sin(2\pi mpb)\sin(2\pi np/b)\tilde{R}(\alpha,\beta;p),
\end{equation}
 with $\tilde{R}(\alpha,\beta;p)$ given in (\ref{Rtil}).
 On the other hand, it follows from the fusion rule that only
 degenerate boundary operators $B_{\beta_{r,s}}$ with
\[
 r=1,3,\cdots, 2m-1,~~~~ s=1,3,\cdots, 2n-1
\]
 can appear on a degenerate Cardy state $\cket{m,n}$.
 To prove the consistency of all these we need further study of
 boundary degenerate operators and degenerate Cardy states.

   Of all the basic structure constants in Liouville theory on a disc,
 it remains to calculate the three-point function of boundary operators.
   The boundary degenerate operator $B_{-b/2}$ and the associated
 KZ equation will be useful also there.

~

   The author thanks the organizer and participants of the Workshop
 on Field Theory at Otaru (Japan) where he initiated this work.
 The work of the author is supported in part by JSPS Research
 Fellowships for Young Scientists. 

\vskip15mm

\begin{center}{\sc References}\end{center}\par

\list{[\arabic{enumi}]}
     {\settowidth\labelwidth{[99]}\leftmargin\labelwidth
      \advance\leftmargin\labelsep\usecounter{enumi}}
\def\newblock{\hskip .11em plus .33em minus .07em}
\sloppy\clubpenalty4000\widowpenalty4000
\sfcode`\.=1000\relax
\let\endthebibliography=\endlist


\bibitem{FZZ} V. Fateev, A. Zamolodchikov and Al. Zamolodchikov,
    {\sl ``Boundary Liouville Field Theory I.
           Boundary State and Boundary Two-point Function''},
    {\tt hep-th/0001012}.
\bibitem{ZZ} A. Zamolodchikov and Al. Zamolodchikov,
    {\sl ``Liouville field theory on a pseudosphere''},
    {\tt hep-th/0101152}.
\bibitem{DO} H. Dorn and H. J. Otto,
    {\sl ``Two and three point functions in Liouville theory''},
    Nucl. Phys. B {\bf 429}, 375 (1994), {\tt hep-th/9403141}.
\bibitem{ZZ2} A. Zamolodchikov and Al. Zamolodchikov,
    {\sl ``Structure constants and conformal bootstrap
           in Liouville field theory''},
    Nucl. Phys. B {\bf 477}, 577 (1996), {\tt hep-th/9506136}.
\bibitem{T} J. Teschner,
    {\sl ``On the Liouville three point function''},
    Phys. Lett. B {\bf 363}, 65 (1995), {\tt hep-th/9507109}.
\bibitem{PT} B. Ponsot and J. Teschner,
    {\sl ``Liouville bootstrap via harmonic analysis
           on a noncompact quantum group''},
    {\tt hep-th/9911110}.
\bibitem{T2} J. Teschner,
    {\sl ``Remarks on Liouville theory with boundary'',}
    {\tt hep-th/0009138}.
\bibitem{T3} J. Teschner,
    {\sl ``Liouville theory revisited'',}
    {\tt hep-th/0104158}.

\end{document}